\documentclass[conference]{IEEEtran}
\IEEEoverridecommandlockouts
\usepackage{cite}
\usepackage{amsmath,amssymb,amsfonts}
\usepackage{mathtools}
\usepackage{graphicx}
\DeclareGraphicsExtensions{.eps,.jpeg,.png,.pdf}
\usepackage{textcomp}
\usepackage{xcolor}

\usepackage{hyperref}
\usepackage{siunitx}
\usepackage{stfloats}
\usepackage{multirow}

\usepackage{amsthm}
\usepackage{pdfpages}
\usepackage{layout}

\def\BibTeX{{\rm B\kern-.05em{\sc i\kern-.025em b}\kern-.08em
    T\kern-.1667em\lower.7ex\hbox{E}\kern-.125emX}}

\newcommand{\Psip}{\psi_{\mathrm{p}}}

\newcommand{\UDC}{U_{\mathrm{DC}}}

\newcommand{\id}{i_\mathrm{d}}

\newcommand{\iq}{i_\mathrm{q}}

\newcommand{\vn}{\boldsymbol{v}_n}
\newcommand{\san}{s_{\mathrm{a},n}}
\newcommand{\sbn}{s_{\mathrm{b},n}}
\newcommand{\scn}{s_{\mathrm{c},n}}

\begin{document}

\title{Data Set Description: Identifying the Physics Behind an Electric Motor -- Data-Driven Learning of the Electrical Behavior (Part I)
\thanks{This work was funded by the German Research Foundation (DFG) under the reference number BO 2535/11-1.}
}

\author{

\IEEEauthorblockN{S{\"o}ren Hanke\IEEEauthorrefmark{1},
									Oliver Wallscheid, 
									Joachim B{\"o}cker}
													
\IEEEauthorblockA{Department of Power Electronics and Electrical Drives, Paderborn University, 33095 Paderborn, Germany,\\\IEEEauthorrefmark{1}E-mail: hanke@lea.upb.de}
}

\maketitle

\begin{abstract}
Two of the most important aspects of electric vehicles are their efficiency or achievable range.
In order to achieve high efficiency and thus a long range, it is essential to avoid over-dimensioning the drive train.
Therefore, the drive train has to be kept as lightweight as possible while at the same time being utilized to the best possible extent.
This can only be achieved if the dynamic behavior of the drive train is accurately known by the controller.
The task of the controller is to achieve a desired torque at the wheels of the car by controlling the currents of the electric motor.
With machine learning modeling techniques, accurate models describing the behavior can be extracted from measurement data and then used by the controller.
For the comparison of the different modeling approaches, a data set consisting of about 40 million data points was recorded at a test bench for electric drive trains.
The data set is published on Kaggle, an online community of data scientists \cite{Dataset}.
\end{abstract}

\section{Preliminary remark}
The description of the data set consists of two parts. Part I (this document) gives a simplified introduction to the system behind the data and explains how to use the data set (\url{https://arxiv.org/abs/2003.07273}) \cite{Description_Part_I}.

Part II explains the system in more details, covers some basic approaches on how to extract models and discusses also a possible way to get a balanced data set where the samples are evenly distributed in a subset used for (deep) machine learning (ML) methods (\url{https://arxiv.org/abs/2003.06268}) \cite{Description_Part_II}.


\section{Introduction}
\begin{figure}[ht]
		\centering
		\includegraphics[]{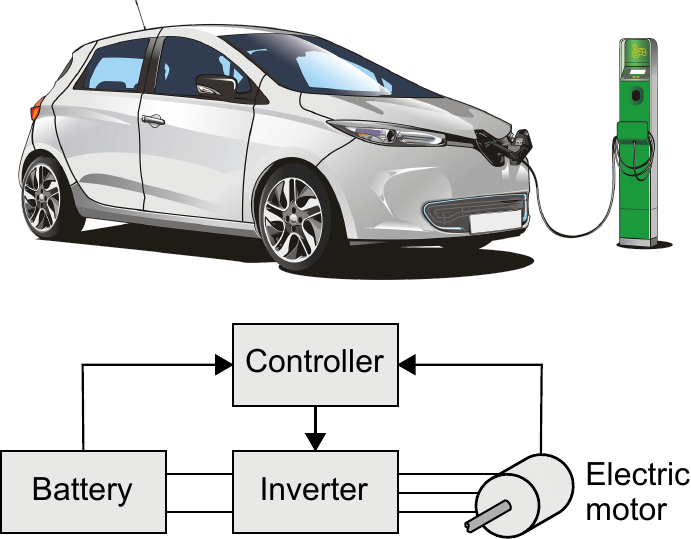}
		\caption{Simplified structure of the drive train in an electric vehicle, \cite{Pixabay}}
		\label{fig:drive_train_structure}
	\end{figure}

The drive train of an electric vehicle consists of a battery, an inverter, an electric motor and a controller (Fig. \ref{fig:drive_train_structure}).

For this dataset, the battery is assumed to be an ideal power supply and, hence, the focus is on the interaction between controller, inverter and motor. The inverter is a power electronic device with switchable semiconductors which converts the electric energy provided by the battery from a two phase DC voltage to a three phase AC voltage with varying amplitude and frequency. This is required, in order to operate the motor at different rotational speeds with a certain torque generated. The motor itself converts electrical to mechanical energy and vice versa. However, at the end the motor is a passive system without any actuators and that is why the controller is only acting on the inverter switching states.  

For a high performance (i.e. accurate, fast and efficient torque control), the controller needs a model that predicts the electromechanical drive's behavior in all operating points sufficiently well.
This means that operating point-dependent effects, such as nonlinearities of the inverter or the magnetic saturation of the motor, must be covered.

To summarize, the objective behind this data set is to predict the electrical behavior of both inverter and motor by one single framework. 

The following two sections first describe the basic operating principle (\ref{Operating_principle}) and then how the data set can be used to obtain models from it including details on choosing the inputs and outputs (\ref{Extracting_models}).

\section{Operating principle}
\label{Operating_principle}

\begin{figure}[ht]
		\centering
		\includegraphics[]{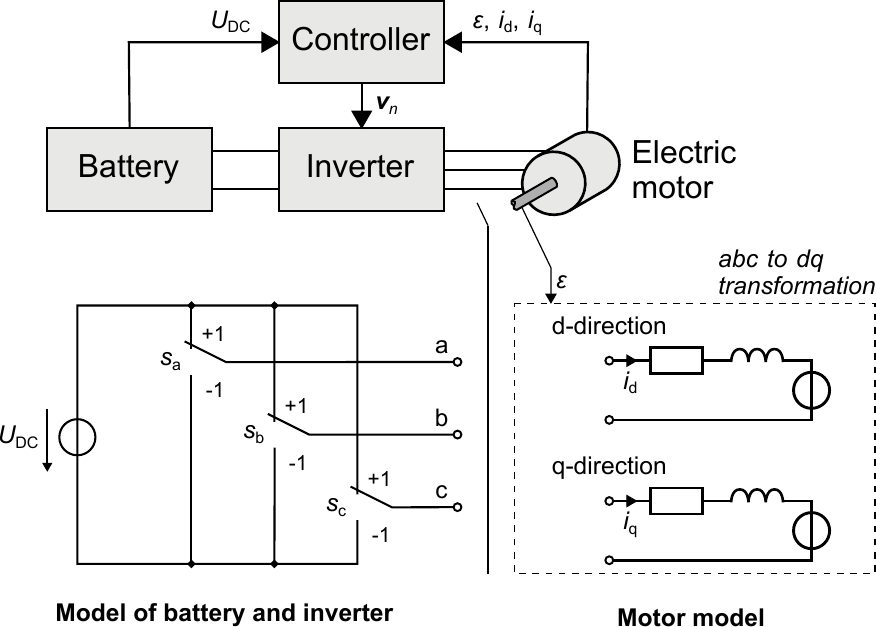}
		\caption{Basic electrical modeling of the drive train}
		\label{fig:ECD_of_drive_train}
	\end{figure}

Fig. \ref{fig:ECD_of_drive_train} shows the components and their basic electrical models in more details.
The battery can be seen as a voltage source having the voltage $\UDC$ and the inverter consists of three switches ($s_\mathrm{a}$, $s_\mathrm{b}$, $s_\mathrm{c}$) each having two possible switching states (+1, -1).
Thus a total of eight different combinations of the switching states yielding different voltages at the inverter terminals (a, b, c) are possible (Tab. \ref{tab:elementary_vectors_}).
These states are also called elementary vectors $\vn$ with $n$ denoting the index of a vector.

If the switching states are modulated / operated in a suitable sequence at a considerable high switching frequency, the inverter outputs a three phase AC voltage with a certain average amplitude and average frequency (fundamental component of the voltage). 

Due to the structure of the inverter, the vectors $\boldsymbol{v}_\mathrm{1}$ and $\boldsymbol{v}_\mathrm{8}$ result in the same voltages at the terminals.
For this reason, it is sufficient to use only vector $\boldsymbol{v}_\mathrm{1}$ in the provided data set.

The motor is modeled in the so called dq-coordinate system which is a typical state transformation in the motor control domain in order to simplify the overall control design.
However, if the momentary rotation angle $\varepsilon$ of the motor is known, the inverter voltages can be transformed to the dq-system an fed into the motor model.

\begin{table}[!ht]
\centering
\caption{Inverter switching states}
\label{tab:elementary_vectors_}
\begin{tabular}{c|ccc|}
\multirow{2}{*}{$n$} & \multicolumn{3}{c|}{$\vn$} \\ \cline{2-4}
                     & $\san$      &  $\sbn$     & $\scn$ \\ \hline
                  1  & -1          & -1          & -1     \\
                  2  & +1          & -1          & -1     \\
                  3  & +1          & +1          & -1     \\
                  4  & -1          & +1          & -1     \\
                  5  & -1          & +1          & +1     \\
                  6  & -1          & -1          & +1     \\
                  7  & +1          & -1          & +1     \\
									8  & +1          & +1          & +1     \\
\end{tabular}
\end{table}	

\begin{figure}[ht]
		\centering
		\includegraphics[]{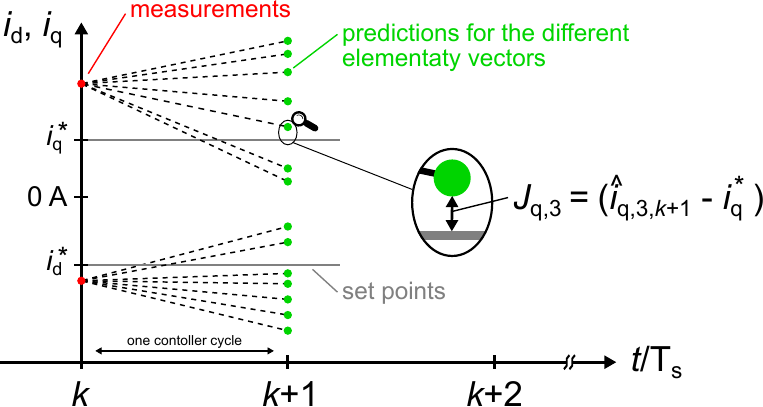}
		\caption{Predictions of the motor currents in d- and q-direction for the possible elementary vectors}
		\label{fig:FCS_MPC_principle_3}
	\end{figure}

\begin{figure}[ht]
		\centering
		\includegraphics[]{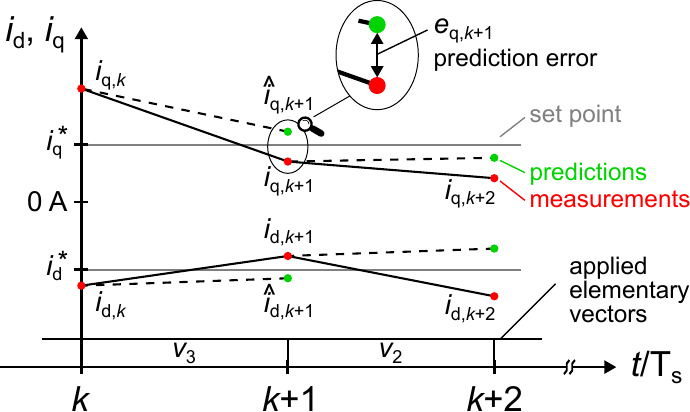}
		\caption{Curve shape of the motor currents with highlighted measurements and predictions}
		\label{fig:FCS_MPC_principle_2}
	\end{figure}

Fig. \ref{fig:FCS_MPC_principle_3} and Fig. \ref{fig:FCS_MPC_principle_2} show, how the controller can regulate the motor currents by selecting appropriate elementary vectors $\vn$.
The motor currents are directly related to the torque but easier to measure and to predict. 
Hence, the basic motor control level is addressing electrical currents flowing through the stator windings. 
As the used controller is based on the model predictive control (MPC) principle, the following steps are carried out recurrently for every controller cycle (explained for the cycle between the time points $k$ and $k+1$):

\begin{enumerate}
	\item Measurement of $\id$, $\iq$, $\varepsilon$ at time point $k$
	\item Prediction of the currents at $k+1$ for the different elementary vectors that can be applied: $\hat{i}_{\mathrm{d},n,k+1}$, $\hat{i}_{\mathrm{q},n,k+1}$
	\item Selection of the elementary vector that brings both currents as close as possible to the set points
	\item Apply the selected elementary vector for one cycle (here between $k$ and $k+1$)
	\item Repeat steps 1-4 for the next cycle beginning at $k+1$\\
\end{enumerate}

For the prediction in step 2, the models that describe the system behavior for the different elementary vectors are used by the controller.
How this models can be extracted from the data set is discussed in section \ref{Extracting_models}.

The selection of the appropriate vector in step 3 is based on the evaluation of the cost function (\ref{eq:cost_function}). 
For each elementary vector the deviation between prediction and set points is evaluated.
Fig. \ref{fig:FCS_MPC_principle_3} shows the deviation for the predicted q-current that would result when choosing vector $\boldsymbol{v}_3$. 
The vector that yields the lowest costs is selected by the MPC.
	
	\begin{equation}
	\begin{split}
			\min_{n} J_n &= J^2_{\mathrm{d},n} + J^2_{\mathrm{q},n} \\
			             &= (\hat{i}_{\mathrm{d},n,k+1} - i^{\ast}_\mathrm{d})^2 + (\hat{i}_{\mathrm{q},n,k+1} - i^{\ast}_\mathrm{q})^2
		\label{eq:cost_function}									
	\end{split}
	\end{equation}
	
However, as shown in Fig. \ref{fig:FCS_MPC_principle_2}, there will be an prediction error due to deviations of the models from the real plant behavior.
But increasing the accuracy of the models will reduce the prediction error and allow MPC to determine the most appropriate vector.

	%
%

\section{How to extract models from the data set}
\label{Extracting_models}

To get accurate models that describe the behavior of the plant (inverter + motor) with a high accuracy, a data-driven approach based on measurements can be used.
The data set provides about 40 million samples that can be used to train for example artificial neural networks or other machine learning models.

A sample in the data set (each row) consists of the measured dq-currents at two consecutive time points (e.g. $k$ and $k+1$), the angle at the earlier of the two time points, and the information about the elementary vector selected in the controller cycle between them ($n_k$) as well as the vector selected in the cycle before ($n_{k-1}$). 
An overview of the included variables is given in Tab. \ref{tab:Variables_contained}.

The rotational speed of the motor for all samples was constant at $n_\mathrm{me}=1000$\,min$^{-1}$ and hence this variable is not part of the data set. 
In the future, an extended data set for varying rotational motor speeds may be added and then the rotational speed would be added to the input space. 
Similar, the motor temperature was nearly constant during all measurements and, therefore, does not need to be considered in the given data set.
  
However, the successive rows or samples in the set do not constitute a time series.

\begin{table}[!ht]
\centering
\caption{Variables contained in the data set}
\label{tab:Variables_contained}
\begin{tabular}{c|c|c|c}
\multirow{1}{*}{Variable} & \multirow{1}{*}{Description} & \multirow{1}{*}{Data type} & \multirow{1}{*}{Classification}\\ \hline \hline
$i_{\mathrm{d},k}$        & measured d-current at $k$ & single & \multirow{5}{*}{inputs} \\
$i_{\mathrm{q},k}$        & measured q-current at $k$ & single &                        \\
$\varepsilon_{k}$         & measured rotational angle at $k$ & single &                        \\
$n_k$                     & element. vector applied at $k$ & integer &                        \\
$n_{k-1}$                 & element. vector applied at $k\!-\!1$ & integer &                        \\ \hline
$i_{\mathrm{d},k+1}$      & measured d-current at $k\!+\!1$ & single & \multirow{2}{*}{targets} \\
$i_{\mathrm{q},k+1}$      & measured q-current at $k\!+\!1$ & single &  \\ \hline
\end{tabular}
\end{table}	

\begin{figure}[ht]
		\centering
		\includegraphics[]{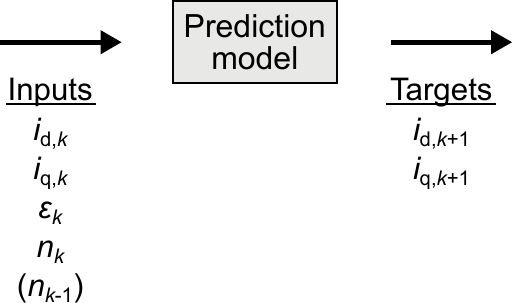}
		\caption{Variant A: Choosing the inputs and targets for a modeling approach using a single model}
		\label{fig:Model_variant_a}
	\end{figure}

\begin{figure}[ht]
		\centering
		\includegraphics[]{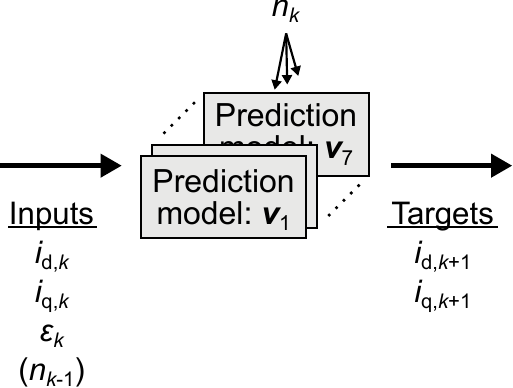}
		\caption{Variant B: Choosing the inputs and targets for a modeling approach using multiple models}
		\label{fig:Model_variant_b}
	\end{figure}

\begin{figure}[ht]
		\centering
		\includegraphics[]{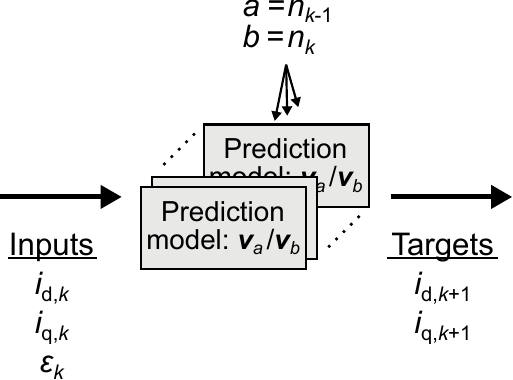}
		\caption{Variant C: Choosing the inputs and targets for a modeling approach using multiple models, with each of them considering the behavior of a particular transition between the former and the next switching vector.}
		\label{fig:Model_variant_c}
	\end{figure}

As a result of the measurements at $k$ and $k+1$, the real behavior of the currents for a given vector is known.
This knowledge can now be used to derive models.
Fig. \ref{fig:Model_variant_a}, Fig.~\ref{fig:Model_variant_b} and Fig.~\ref{fig:Model_variant_c} show three variants on how to define the inputs and targets for a machine learning modeling.

Variant A uses a single model that aims to cover the behavior of all elementary vectors.
In addition to the known values at time $k$, the index of the vector to be analyzed $n_k$ is input to the model.
The information about which vector was used in the interval before ($n_{k-1}$) can also be an input. 
This might be helpful to consider more detailed effects like the inverter-deadtime or the interlocking time, as they appear when switching between elementary vectors.
Targets are the currents at the end of the controller cycle.

Nevertheless, it is also possible to extract multiple models each covering the specific behavior of one of the used vectors (variant B).
The index of the vector to be analyzed ($n_k$) is then used by the controller to switch between the models.

To account for the vector which was used in the interval before ($n_{k-1}$), it is also possible extract $7^2=49$ models.
In this variant C, each model covers the behavior of a particular transition between the former vector $n_{k-1}$ and the next vector $n_{k}$ as well as the behavior of the currents with vector $n_{k}$ applied during the next cycle.

The accuracy of a model can be evaluated according to the cost function (\ref{eq:cost_function_model}).

	\begin{equation}
	\begin{split}
	J_\mathrm{model} = \sqrt{ \frac{1}{m} \sum^{m}_{j=1} \left( e^2_{\mathrm{d},k+1,j} + e^2_{\mathrm{q},k+1,j} \right) }\\
	\mathrm{with}\quad e_{\mathrm{d},k+1,j} = \hat{i}_{\mathrm{d},k+1,j} - i_{\mathrm{d},k+1,j}\\
	\quad \quad e_{\mathrm{q},k+1,j} = \hat{i}_{\mathrm{q},k+1,j}  - i_{\mathrm{q},k+1,j}\\
	\label{eq:cost_function_model}
	\end{split}
	\end{equation}

Considering $e_{\mathrm{d},k+1,j}$ and $e_{\mathrm{q},k+1,j}$ as elements of an error vector for each sample $j$, the cost function represents the root mean square of this error vector (RMSE) regarding the $m$ samples used for the training of the respective model. The error vector elements are the deviation between the targeted outputs ($i_{\mathrm{d},k+1}$, $i_{\mathrm{q},k+1}$) and the predictions of the model ($\hat{i}_{\mathrm{d},k+1}$, $\hat{i}_{\mathrm{q},k+1}$) for the d- and the q-current.\\

\noindent \textbf{Please note:}\\
For sake of simplicity, $\UDC$ is considered as ideally constant in this contribution.
Moreover, the rotational speed $n_\mathrm{me}$ and the motor temperature are kept constant, too.
It is planed to extend the data set to variations of this three variables in the future.
However, the presented data-driven modeling ideas can be directly extended to consider these varying operation conditions by extending the input space with this additional features. \newline

%
%
%
%
%
%

%

\begin{figure}[ht]
		\centering
		\includegraphics[width=8.0cm]{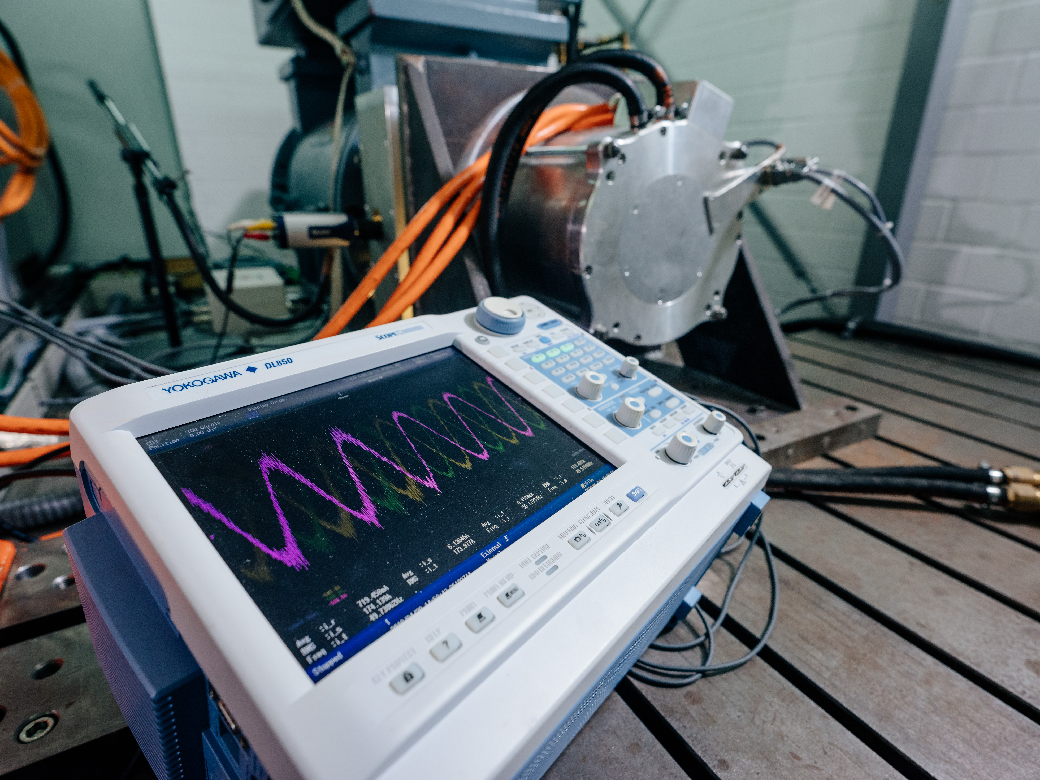}
		\caption{Test bench with the used PMSM in the background}
		\label{fig:test_bench}
	\end{figure}

%
\noindent \textbf{Devices under test:}\\
The drive system under test consists of an interior magnet permanent magnet synchronous motor (IPMSM) of \SI{57}{\kW} and a 2-level IGBT inverter. 
The most important test bench parameters are summarized in Tab. \ref{tab:param}.
Fig. \ref{fig:test_bench} shows the test bench with the transient recorder in the front and the used motor in the background.

Are more detailed list of the available equipment can be found in Part II of the description.

\begin{table}[ht]
				\caption{Test bench parameters}
			  \label{tab:param}
				\centering
				\begin{tabular}{l|c|c}
				\hline
				\textbf{IPMSM} & \multicolumn{2}{c}{Brusa HSM16.17.12-C01} \\
				Stator resistance         & $R_\mathrm{s}$    & 18\,m$\Omega$ 						\\ 
				Inductance in d-direction & $L_\mathrm{d}$    & \SI{370}{\micro\henry}    \\ 
				Inductance in q-direction & $L_\mathrm{q}$    & \SI{1200}{\micro\henry}   \\
				Permanent magnet flux     & $\Psip$           & 66\,mV\,s                 \\
				Pole pair number          & $p$               & 3                         \\ \hline
				\textbf{Inverter} & \multicolumn{2}{c}{3$\times$SKiiP 1242GB120-4D} \\
				Typology                  & \multicolumn{2}{c}{voltage source inverter} \\
				        									& \multicolumn{2}{c}{2-level, IGBT} \\          \hline
				\textbf{Controller hardware} & \multicolumn{2}{c}{dSPACE} \\
				Processor board              & \multicolumn{2}{l}{DS1006MC, 4 cores, 2.8\,GHz} \\ \hline
				\textbf{Measurement devices} & \multicolumn{2}{c}{}  \\
				Transient recorder & \multicolumn{2}{c}{Yokogawa\ \ \ DL850} \\
				Power analyzer     & \multicolumn{2}{c}{Yokogawa WT3000} \\
				Current probes (zero-flux transducers) & \multicolumn{2}{l}{3$\times$Yokogawa, 500\,A, 2\,MHz}  \\	 
				Torque sensor 												 & \multicolumn{2}{l}{\ \ \ \ \, HBM, T10FS, 2\,kN\,m} \\  \hline 
				\end{tabular}
		\end{table}	

\noindent \textbf{Link to the uploaded data set:}\\
The data set is published on Kaggle, an online community of data scientists: \url{https://www.kaggle.com/hankelea/system-identification-of-an-electric-motor}

\bibliography{refs}

\end{document}